\documentclass[usenames,dvipsnames,journal]{IEEEtran}

\usepackage{pgfplots}
\usepackage{lipsum}
\usepackage{multicol}
\usepackage{cite}
\usepackage{textcomp}
\def\BibTeX{{\rm B\kern-.05em{\sc i\kern-.025em b}\kern-.08em
    T\kern-.1667em\lower.7ex\hbox{E}\kern-.125emX}}
\usepackage [english]{babel}

\usepackage{color, colortbl}
\usepackage{amsmath}
\usepackage{algorithmic}
\usepackage[english]{babel}
\usepackage[utf8]{inputenc}
\usepackage{algorithm}
\usepackage{array}
  \usepackage[caption=false,font=footnotesize]{subfig}
\usepackage{caption}

\usepackage{fixltx2e}
\usepackage{stfloats}
\usepackage{url}
\usepackage{eqparbox}
\usepackage{arydshln}
\begin{document}
\captionsetup[figure]{labelfont={bf},labelformat={default},labelsep=period,name={Fig.}}
\captionsetup[table]{justification=centering, labelsep=newline}
\title{\title{AutoFoley: Artificial Synthesis of Synchronized Sound Tracks for Silent Videos with Deep Learning}}

\author{Sanchita~Ghose,~\IEEEmembership{Student Member,~IEEE} and ~John~J.~Prevost*,~\IEEEmembership{Member,~IEEE}
\thanks{S. Ghose (sanchita.ghose@my.utsa.edu) and J. J. Prevost (jeff.prevost@utsa.edu) are with the Department of Electrical and Computer Engineering, The University of Texas at San Antonio, One UTSA Circle, San Antonio, TX 78249 USA.\newline
*- Indicates the Corresponding Author\newline
This research is supported by the Open Cloud Institute at UTSA.
}}

%

\maketitle

\begin{abstract}
In movie productions the Foley Artist is responsible for creating an overlay sound track that helps the movie come alive for the audience. This requires the artist to first identify the sounds that will enhance the experience for the listener thereby reinforcing the Director’s intention for a given scene. The artist must decide what artificial sound captures the essence of both the sound and action depicted in the scene. In this paper we present AutoFoley, a fully-automated deep-learning tool that can be used to synthesize a representative audio track for videos. AutoFoley can be used in applications where there is either no corresponding audio file associated with the video, or in cases where there is a need to identify critical scenarios and provide a synthesized, reinforced sound track. An important performance criterion of the synthesized sound track is to be time-synchronized with the input video, which provides for a realistic and believable portrayal of the synthesized sound. Unlike existing sound prediction and generation architectures, our algorithm is capable of precise recognition of actions as well as inter-frame relations in fast moving video clips by incorporating an interpolation technique and Temporal Relational Networks (TRN). We employ a robust multi-scale Recurrent Neural Network (RNN) associated with a Convolutional Neural Network (CNN) for a better understanding of the intricate input-to-output associations over time. To evaluate AutoFoley, we create and introduce a large-scale audio-video dataset containing a variety of sounds frequently used as Foley-effects in movies. Our experiments show that the synthesized sounds are realistically portrayed with accurate temporal synchronization of the associated visual inputs. Human qualitative testing of AutoFoley show over 73\% of the test subjects considered the generated sound track as original, which is a noteworthy improvement in cross-modal research in sound synthesis.
\end{abstract}

\begin{IEEEkeywords}
Foley, convolutional neural network, recurrent network,  interpolation, multi-scale temporal relational network, cross-modal learning, sound synthesis.
\end{IEEEkeywords}

%
\IEEEpeerreviewmaketitle

\section{Introduction}
\IEEEPARstart{A}{dding} sound effects in post production using the art of Foley has been an intricate part of movie and television soundtracks since the 1930s. The technique is named after Jack Foley, a sound editor at Universal Studios. Mr. Foley was the first to make sound effects for live radio broadcasts with the tools and items he had around him. Now almost every motion picture and television show contain Foley tracks. Movies would seem hollow and distant without the controlled layer of a realistic Foley soundtrack.

To construct the augmented sound, the Foley artist uses special sound stages (Fig.1) surrounded by a variety of props such as car fenders, chairs, plates, glasses as well as post production sound studios to record the sound effects without the ambient background sounds. This requires electronics such as monitors, camcorders, mikes, manual record volume controller, and an external audio mixer. Foley artists have to closely observe the screen while performing diverse movements (e.g. breaking objects, running forcefully on rough surfaces, pushing each other, scrubbing different props) to ensure their sound effects are appropriate (Fig. 2). The process of Foley sound synthesis therefore adds significant time and cost to the creation of a motion picture.
Furthermore, the process of artificially synthesizing sounds synchronized to video multi-media streams is a problem that exists in realms other than that of the Motion Picture industry. Research has shown that the end-user experience of multimedia in general is enhanced when more than one source of sensory input is synchronized into the multimedia stream \cite{yuan2015}.

\graphicspath{ {Figure} }
\begin{figure}[]
\includegraphics[width=87mm]{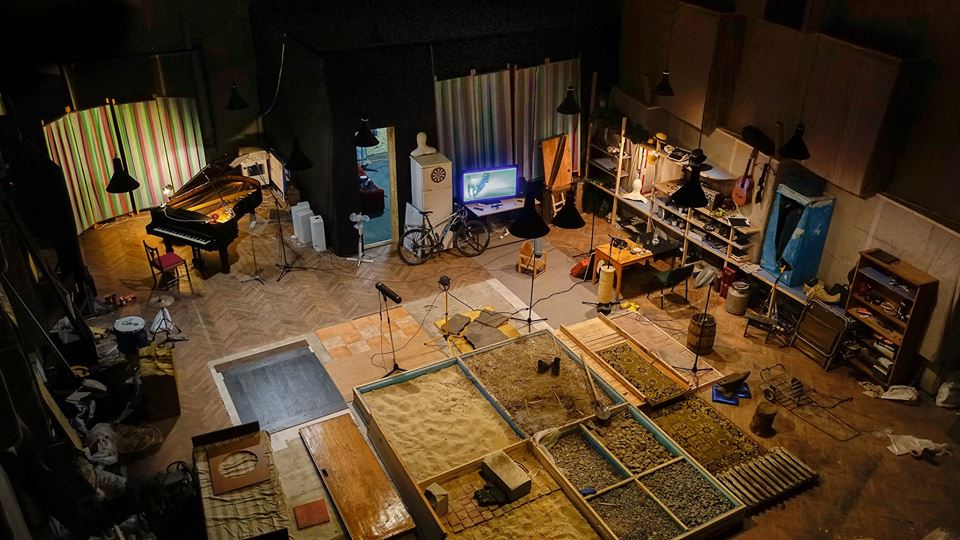}
\caption{Foley Recording Studio \cite{foley}.}
\label{Fig:1}
\end{figure}

\graphicspath{ {Figure} }
\begin{figure}[]
\includegraphics[width=87mm]{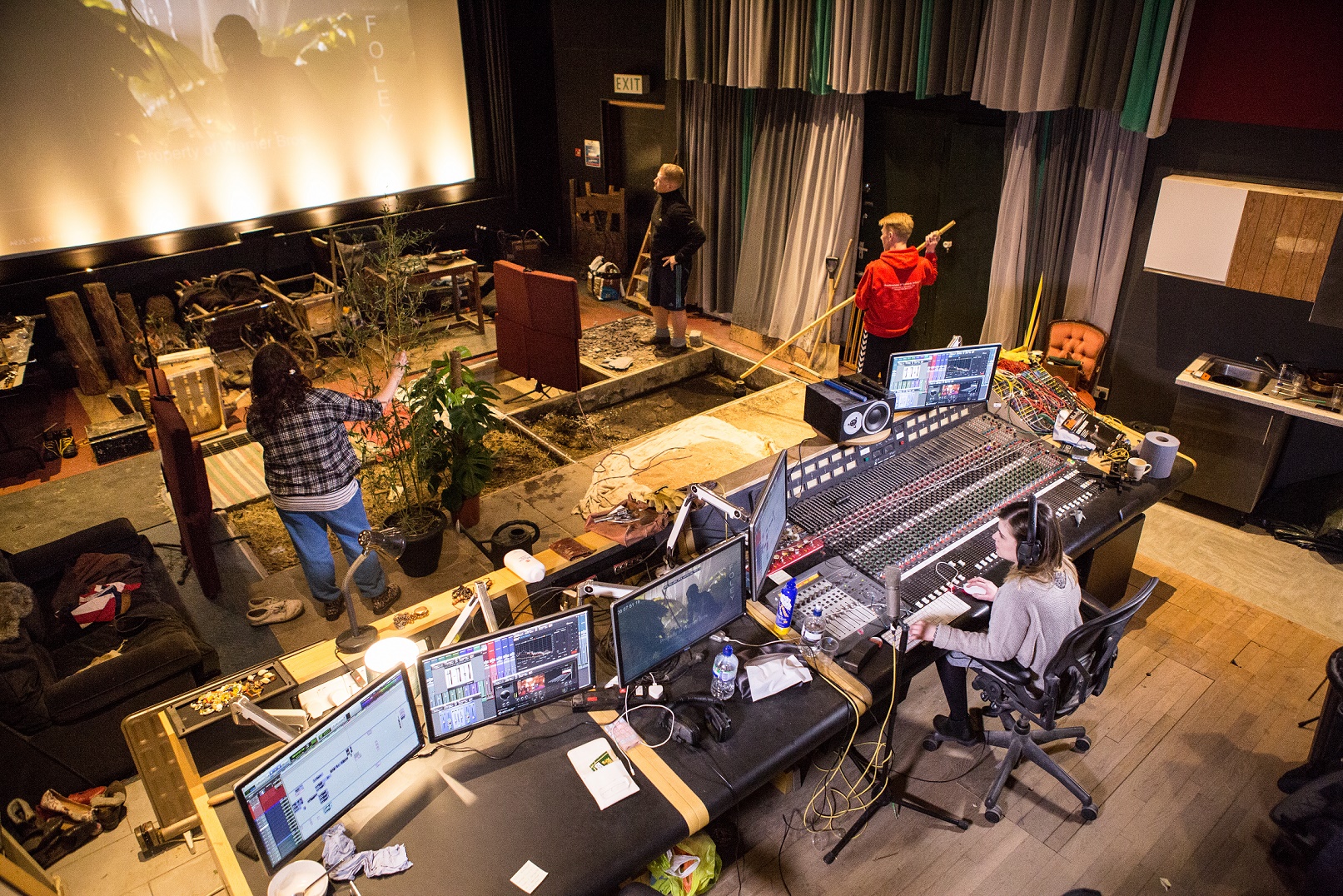}
\caption{Foley Recording for movie at Pinewood Studio (https://www.avid.com/).}
\end{figure}
 
We generate the augmented sounds for injection into a synchronized video file by first utilizing a deep neural network. To train our network, we find motivation from the multi-modal coherence ability of the human brain to synchronize audio and visual signals. Recent works in \cite{owens2018audio, owens2016visually,owens2018learning, zhou2017visual,takahashi2017aenet} explored the relationship between auditory and visual modalities through computational models as way of localizing and separating sound source, material and action recognition, generating natural sounds and learning audio features for video analysis. In this paper, we propose a deep sound synthesis network for the first time that performs as an automatic Foley, generating augmented and enhanced sound effects as an overlay on video files that may or may not have associated sound files. 

\graphicspath{ {Figure} }
\begin{figure}[t]
\includegraphics[width=85mm]{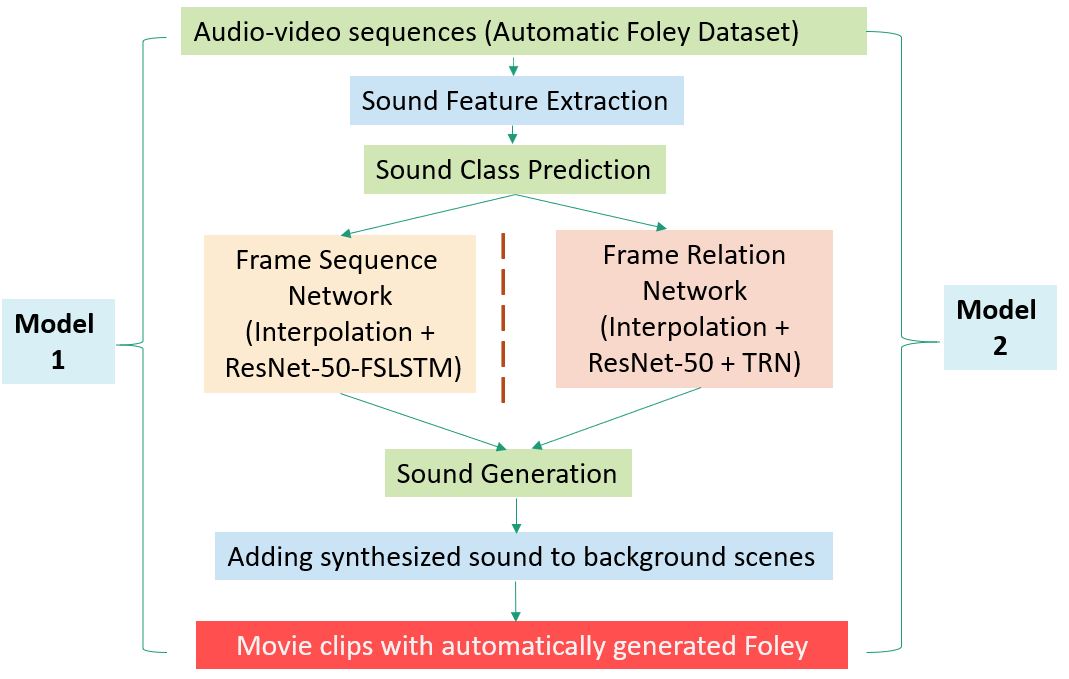}
\caption{Automatic Foley Generation Model: Model 1 architecture (Frame Sequence Network+sound synthesis) shown in left; Model 2 architecture (Frame Relation Network+sound synthesis) shown in right.}
\end{figure}

In our proposed system (Fig.3), we present two distinct models, using different audio-video regression approaches, followed by a sound synthesis architecture. The first model maps the auditory features to interpolated video frames associated with a combined multiscale deep convolutional-recurrent neural network (ResNet-FSLSTM). Instead of replicating the same video frame multiple times in order to map with a single sound sample (\cite{owens2016visually, zhou2017visual}), we use the interpolation technique \cite{liu2017video} to leverage intermediate video frames to obtain smooth motion information of each video. Again, sound class prediction executing a CNN-FSLSTM architecture provides better understanding of the intricate transition functions over varying time scales. In the second model, we predict the sound category only from the representative frames of each video applying a Temporal Relation Network (TRN) \cite{zhou2018temporal}. This enables the system to learn relational reasoning between two frames of different time periods, identifying the anticipated action present in the video. 

More specifically, our first model is designed to learn the audio-visual relation for fast moving movie clips, whereas the second model is designed for early recognition of activity using limited video frames. For sound generation, we use the Inverse Short Time Fourier Transform (ISTFT) method \cite{welch1967use}, which has less computational complexity than a parametric synthesis approach \cite{mcdermott2011sound, slaney1995pattern} or an example based synthesis method \cite{owens2016visually}.

To enable learning, we introduce a dataset that contains video scenes associated with audio tracks that are very common in movie clips. Existing works \cite{owens2016visually, zhou2017visual} show sound synthesis for given input of videos of hitting different objects, or of videos collected in the wild. In our work, we attempt to synthesize sound from video clips that are similar to the manually generated Foley sound created by the Foley artists. The advancements introduced by our paper are: 
\begin{itemize}
  \item We take the first step toward automatic Foley generation in silent video clip using deep neural networks, taking into consideration the "movie sound effects" domain, where the desired sounds are highly varied and have a clear temporal onset.
  \item We introduce a dataset explicitly for future Foley synthesis application. We carefully prepare the dataset considering the relevance with movie events without background noise.
  \item We present efficient prediction architectures for realistic and synchronous sound synthesis from visual scenes.
  \item We show Temporal Relation Networks can be exploited in video-to-sound prediction task.
  \item For the performance analysis of generated sounds, we perform qualitative, numerical experiments and conduct a human survey on our generated sounds. We also present a detail ablation study of our proposed methods to show the usefulness of each component.
\end{itemize}

The rest of the paper is organized as follows, in section II, we present a brief review of related works. In section III, we describe our methodology (audio and video pre-processing steps, training procedure with deep neural network, sound synthesis process) and the full algorithm in detail. Finally in section IV and V, we conclude with the explanation of our AutoFoley dataset resembling video clips with Foley sound tracks, parameter details for experiments, along with qualitative, numerical, ablation and human experiments to evaluate the proposed models and the generated results.


\section{Related work}
\subsection{Automatic Sound Effect}
Previous work in \cite{van2001foleyautomatic} proposes dynamic simulation and user interaction to produce sound effects automatically from 3D models. They use contact forces modeled at audio rates to drive modal models. Work in \cite{hahn1998integrating} presents a number of similar algorithms for contact sound synthesis. In our work, we use a deep neural network to predict sound in silent video clips of movie scenes, then synthesize Foley sound automatically from the predicted features. This represents a novel methodology for automatic Foley sound generation application. In this paper, we are proposing for the first time automatic Foley sound synthesis using deep neural networks.
\subsection{Audio-Visual Correlation}
Our approach is extensively inspired by research in \cite{owens2016visually, zhou2017visual, castrejon2016learning, ngiam2011multimodal, huang2013audio, chen2017deep, zhangvisually}, and on learning the cross-modal modality relationship from \cite{baltruvsaitis2019multimodal}. Among these, \cite{owens2016visually} deploy a deep learning algorithm to predict impact sounds based on different materials and physical actions. Their work is focused on material recognition. Research works in \cite{owens2016visually} and \cite{chen2017deep} aim to generate sound directly from input video frames rather than the natural synchronization property between audio and video. Earlier works in  \cite{arandjelovic2017look, owens2016ambient, aytar2016soundnet} 
train their networks with unlabeled video data utilizing the correlation property of video and sound. An unsupervised two stream neural network is proposed in \cite{arandjelovic2017look} that takes one audio and video frames as inputs to identify whether there is coherence or not. Work in \cite{aytar2016soundnet} presents efficient sound representations with a deep convolutional network inspired by teacher-student models \cite{ba2014deep, gupta2016cross} to learn under visual supervision. 
In the work \cite{niwa2018efficient} the minimum spatial resolution of human auditory localization from visual information maintaining a decent alignment is explored. In this paper, we exploit the cross modal relationship between audio and video signal to automatically generate Foley tracks in silent videos.

\subsection{Regression Methods}
Sound regression to videos for the material recognition task is performed in  \cite{owens2016ambient}, a combined CNN-LSTM structure is proposed to map sound (represented by cochleagram \cite{muthusamy1990speaker}) from input video frames. CNN is used for predicting statistical summary of the
sound from a video frame in \cite{owens2016ambient} and \cite{owens2018learning}. A deep CNN architecture, called AENet is proposed in \cite{takahashi2017aenet} to identify long-time audio event in video and  learn the audio features for video analysis. On the other hand, \cite{chen2017deep} pose a GAN structure for the purpose of sound generation of musical instruments influced on visual modality. The SampleRNN \cite{karpathy2014large} structure is deployed by \cite{zhou2017visual} to synthesis visually indicated raw sound wave from video features. In our work, we apply spectrogram features and interpolated image features to train an optimized CNN+LSTM network followed by ISTFT algorithm for the sound regression task. 
\subsection{Action Recognition and Sound Source Detection}
Human capability of localizing sound source explained in \cite{gaver1993world, kingma2014adam, majdak20103, shelton1980influence, bolia1999aurally, perrott1996aurally} describes how humans can adopt the correlation between sound and visual domains in an event-centric way, and learn about different materials of objects and various events from sound by day-to-day listening. Similarly \cite{senocak2018learning} proposes a learning network for the sound source using an attention mechanism \cite{xu2015show}. Recently, action recognition by extracting spatial-temporal features from a video exploiting semantic guided modules (SGMs) is performed in \cite{yu2019weakly}. In \cite{wang2016visualizing} an automatic video sound recognition and visualization framework is proposed where nonverbal sounds in a video are automatically converted into animated sound words and are placed close to the sound source of that video for visualization. Previous research in \cite{majdak20103, bolia1999aurally, perrott1996aurally} provide advanced approaches on localizing sound source against visual data in 3D space. Our Automatic Foley system is not only able to recognize the action in a visual scene, but also capable to add realistic sounds to that video.  
\subsection{Comparison with state-of-the-art \cite{zhou2017visual}}
There are some major differences between \cite{zhou2017visual} and this work. First, \cite{zhou2017visual} introduced a sound generation task from videos in the wild, but our goal is to build the automatic Foley generation network that learns from multi-modal coherence ability. Second, they presented three variants to encode the visual data and combined them with the sound generation network using a video encoder and sound generator structure. Conversely, we focus on sound classes to adopt key variations, where each category of action produces sounds belonging to a particular class, then exploit class predictions to synthesize more realistic and finely tuned sounds. Therefore, we present two distinct audio-video regression models that predict sound class and transform the predicted sound class’s base sample (average sound clip of same class) to visually coherent sound. Our first model is designed to learn the audio-visual relation for fast moving movie clips whereas the second model learns early recognition of activity from limited video frames. In addition, we apply video interpolation technique over the flow based method used in \cite{zhou2017visual} for better interpretation of image features from video frames. Lastly, large number of videos derived from AudioSet are accumulated in the VEGAS dataset \cite{zhou2017visual} but this is unable to match our requirements as most of them contain either background sound or human speech. In this paper, we create our Automatic Foley dataset entirely focused on Foley sound tracks of 12 different movie events. Finally, we overcome the limitations faced in sound synthesis with SampleRNN \cite{zhou2017visual} with a simpler sound generation process (described in detail in methodology and ablation analysis).

\section{Methodology}
The methodology consists of three major steps: i) Sound feature extraction ii) Sound class prediction from video frames and iii) Sound synthesis (explained in the following subsections). The architecture of automatic foley track generation from visual inputs is presented in Fig. 4.

\graphicspath{ {Figure} }
\begin{figure}[]
\includegraphics[width=90mm]{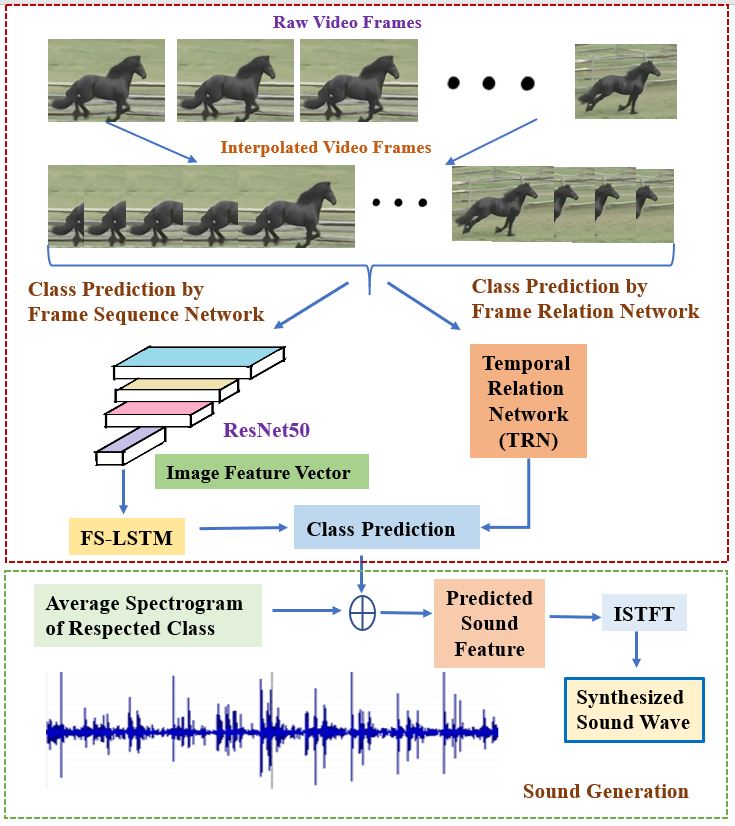}
\caption{Automatic Foley Generation Architecture: the red and green border lines are showing detailed stages of sound prediction and generation steps to synthesize automatic Foley tracks from visual scenes.}
\end{figure}

\subsection{Sound Feature Extraction}
We first compute the features of all the audio files using spectrogram analysis; a visual way of representing the strength of a signal over time at different frequencies present \cite{zhang2018learning}. Human hearing is based on a form of real-time spectrogram encoded by the cochlea of the inner ear, we convert the audio signal into a 2D representation (spectrogram) for extracting the audio feature. Our spectrograms provide an intensity plot (in dB) of the Short-Time Fourier Transform (STFT) magnitude of the windowed data segments along with the window width $w$ 
by computing as follows : 
\begin{equation}
 spectrogram(t,w)=|STFT(t,w)|^2  
 \end{equation}
The windows are usually allowed to overlap by 25\%-50\% in time. We use 950 frames since our sampling frequency is 44100 Hz. A Hanning window is used to compute STFT, and each audio feature vector ${S_n(t)}$ has a width of 1 and a height of 129. 
    
 \graphicspath{ {Figure} }
 \begin{figure}[]
 \includegraphics[width=87mm]{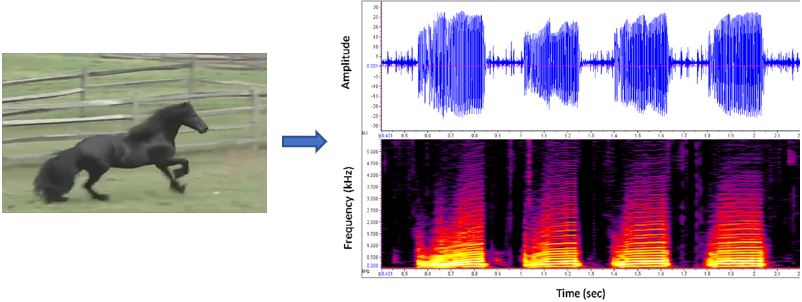}
 \caption{Oscillogram and Spectrogram Plot of a horse running video. The upper blue-colored oscillogram plot presents the waveform and amplitude of the sound over time (sec). The lower figure is the spectrogram plot (frequency (kHz) vs. time (sec) plot) showing the change of the non-stationary signal’s frequency content over time.}
 \end{figure}

 In spectrogram plot (Fig.5), the intensity of the color represents the amount of energy present in each frequency. The brighter the color, the more energy is present in the audio at that frequency.

\subsection{Sound Class Prediction from Video}
We present two different approaches for predicting the sound class from input video frames: i) Frame-Sequence Network (interpolation technique followed by combined convolutional neural network (CNN) and Fast-Slow LSTM (FS-LSTM) network and ii) Frame-Relation Network (combination of CNN and temporal relational network (TRN)).

\subsubsection{Approach 1: Frame-Sequence Network}
In this approach, we increase the video frame rate capturing detail motion information of video frames by using interpolation technique. Next, we extract the image features from each interpolated video frame by applying CNN (ResNet-50). Finally, we predict the sound class associated with the video clip by recurrent network (FS-LSTM) using the image features obtained from the bottleneck convolutional layer of the residual network.

\paragraph{Video Preprocessing using Interpolation Technique}
To map image features with audio features, we face the problem of equalizing the video frame numbers with audio sample numbers. In our dataset we find, the total number of audio samples is about 13 times higher than the total video frame number. In earlier works \cite{owens2016visually} each image feature vector was replicated k times (where k is the ratio of audio and video sampling rates) to compensate for the difference between the two sampling rates. However, this replication technique sometime fails to store the precise motion information of videos correctly, thus we lose some action relationships between consecutive frames. We solve this problem exploiting the technique of generating intermediate video frames by using interpolation algorithm during the video pre-processing step before moving towards the image feature extraction step. 
Video frame interpolation technique can estimate dense motion, typically optical flow between two input frames, then interpolate one or more intermediate frames guided by the motion \cite{niklaus2017video}. Because of this interpolation, we successfully increase the video's frame rate as required without changing the video length and losing the inter-frame motion information. 

\paragraph{Generating Image Feature Vector Using CNN}
We aim to compute image feature vectors with the ResNet-50 convolutional neural network (CNN) model. For this, first we read a sequence of interpolated video frames $I_1, I_2,...,I_n$ from training videos as visual inputs excluding corresponding sound tracks. To avoid the complexity found in the two stream approach \cite{donahue2015long} while obtaining accurate flow estimates in case of fast, non-rigid motion videos, we create space-time images ($SP_1, SP_2,...,SP_n$)  for each frame by joining three gray-scaled images of the previous, current, and next frames (Fig. 6). 

 \graphicspath{ {Figure} }
 \begin{figure}[]
 \tolerance 9999
 \includegraphics[width=83mm]{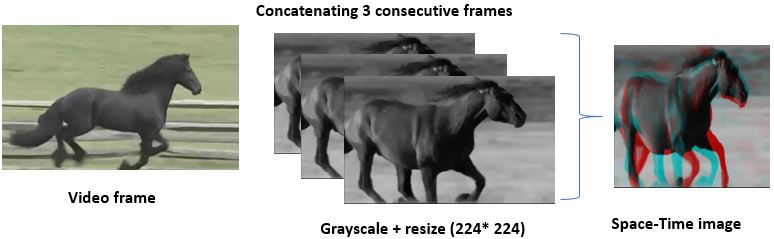}
 \caption{Space-time Image Generation: Concatenation of three consecutive $224 \times 224$ resized grayscale frames of horse racing video after applying interpolation technique.}
 \end{figure}

Finally, we compute the input feature vector $V_t$ for each frame $t$, by concatenating the CNN features for the space-time image $(SP_t)$ at frame $t$ and the RGB image from the first frame of each video : 
\begin{equation}
V_t = [f(SP_t); f(I_1)]
\end{equation}
Here, $f$ represents the CNN features obtained from the output of $conv5$ layer of the ResNet-50 architecture using a Keras pretrained model for ResNet-50 \cite{He_2016_CVPR}. Thus, each of our image features contain motion information (obtained from the space-time images) and color information (obtained from the first RGB frame). The entire image feature extraction process is shown in Fig.7.

 \graphicspath{ {Figure} }
 \begin{figure}[]
 \includegraphics[width=90mm]{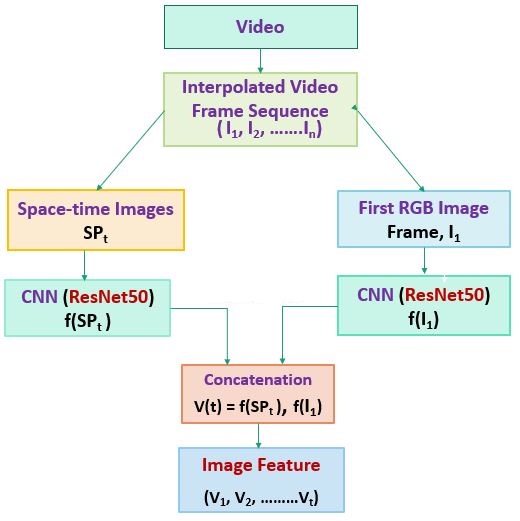}
 \caption{Image Feature Extraction in Frame-Sequence Network applying Convolutional Neural Network (CNN).}
 \end{figure}
 
\paragraph{Sound Class Prediction Using FS-LSTM}

We use the image features ($V_t$) found from our CNN as input to a special recurrent neural network (RNN) named as Fast-Slow LSTM (FS-LSTM), proposed in \cite{mujika2017fast}. Here, LSTM cells act as building units to map the video frames to audio frames and compute the predicted sound features ($\vec{s'_t}$) at each time instance. The FS-LSTM network has two hierarchical layers. The lower layer contains $n$ number of sequential LSTM cells ($L_1$,$L_2$,...,$L_N$) considered Fast cells, whereas the upper layer consists of only one LSTM cell ($U$) considered the Slow cell. $U$ takes input from $L_1$ and gives its state to $L_2$. In the lower layer, the image feature vector $V_t$ is fed to the first LSTM cell ($L_1$) as input and the final predicted sound class matrix is computed as the output of the last LSTM cell ($L_N$). For arbitrary LSTM cells $L_1$, $L_2$,...,$L_N$ and $U$, the FS-LSTM architecture (Fig.8) can be expressed in the following set of equations:
\begin{equation}
\begin{split}
&H{^{L_1}_t} =\beta^{L_1}(H{^{L_n}_{t-1}},V_t)\\
&H{^U_t} =\beta^U(H{^U_{t-1}},H{^{L_1}_t})\\
&H{^{L_2}_t} =\beta^{L_2}(H{^{L_1}_t},H{^U_t})\\
&H{^{L_i}_t} =\beta^{L_i}(H^{L_{i-1}}_t) ; 3\leq i \leq N\\
\end{split}
\end{equation}

Here, $\beta$ represents the update function of a single LSTM cell. The encoding process is performed by revising the value of hidden vector ($H_t$) with updated image feature vector $(V_t)$. Finally, the output sound class prediction matrix ($s_{c1}$) is calculated from affine transformation of $H{^{L_N}_t}$ :
\begin{equation}
s_{c1} = WH{^{L_N}_t} + B
\end{equation}

 \graphicspath{ {Figure} }
 \begin{figure}[]
 \begin{center}
 \includegraphics[scale=1]{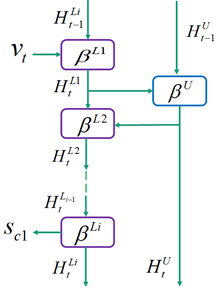}
 \end{center}
 \caption{Fast-Slow LSTM network with i Fast cells.}
 \end{figure}

\subsubsection{Approach 2: Frame-Relation Network}

We aim to capture the detail transformations and actions of the objects, present in the movie scenes more accurately with less computation time. So, to learn the model about the complex behaviors of a visual scene with less amount of video frames, we apply an interpretable network combining CNN and Multi-Scale Temporal Relation Network (TRN), proposed in \cite{zhou2018temporal}. The network compiles the temporal relations among the frames at different time scales by using the following equation:

\begin{equation}
MR_Q(v) = R_2(v) + R_3(v) +...+R_Q(v)
\end{equation}

Here, $R_2$, $R_3$,...,$R_Q$ are the relation functions that capture the temporal relations between 2, 3,...,Q number of ordered frames of video $v$ with respectively. We compute the relation over time among up to 8 frames by setting $Q$ equals to 8. If the video $v$ contains $r$ number of selected frames as $v =[{f_1, f_2,...,f_r}]$, we can define relation functions as follows:

\begin{equation}
\begin{split}
&R_2(v) = h_\phi(\sum_{j<k}g_\theta(f_j,f_k))\\
&R_3(v) = h'_\phi(\sum_{j<k<l}g'_\theta(f_j,f_k,f_l))\\
\end{split}
\end{equation}

where $f_j$, $f_k$, $f_l$ are denoting the $j^{th}$, $k^{th}$, $l^{th}$ frame of the video. $h_\phi$
and $g_\theta$ functions are derived from Multilayer Perceptrons (MLP) layers to combine features of the uniformly sorted video frames and both are unique for every relation function $R(v)$. The same ResNet-50 pretrained model (discussed in approach 1) is used as base CNN to extract visual features from the videos. Next, we fuse several random features together to get their frame relations and then finally feed them to TRN modules to obtain the sound class $s_{c2}$. The multi-scale TRN architecture is illustrated for 2-frame, 3-frame and 4-frame relations is shown in Fig 9.  

\graphicspath{ {Figure} }
\begin{figure}[]
\begin{center}
\includegraphics[width=90mm]{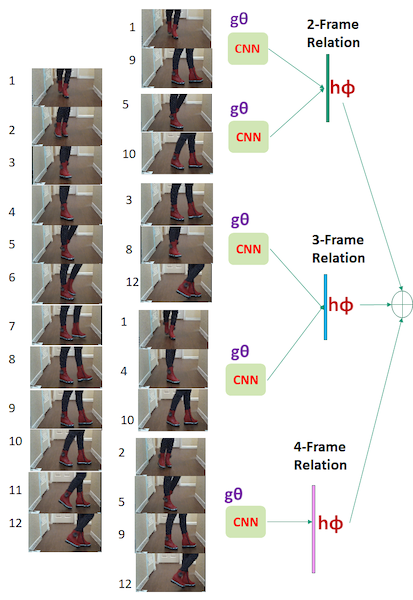}
\end{center}
\caption{Multi-Scale Temporal Relation Network (TRN): the network is learning temporal dependency among representative frames of a visual scene. }
\end{figure}

\subsection{Sound Synthesis}

We apply the same sound synthesis method on both of the sound class prediction approaches. We take the average of all spectrograms of each sound class in our training set, that we combine with the predicted sound class matrix $s_c$ computed from the frame sequence and the frame relation network separately. For the $K^{th}$ sound class, we find the predicted sound feature, $s'(K)$ from its average spectrogram ($A_K$):
\begin{equation}
s'_t(K) = s_c + A_K
\end{equation}
We reduce the difference between the actual the and predicted sound feature for every timestep, calculating the robust loss $L$ through prediction of the square root of the sound features:
\begin{equation}
L(\gamma) = log(\alpha +\gamma^2)
\end{equation}
\begin{equation}
E(\vec{s'_t}) = \sum_{t=1}^T L(||\vec{s_t} - \vec{s'_t}||_2)
\end{equation}

For sound synthesis from predicted sound feature vectors $(\vec{s'_t})$, we perform the Inverse Short Time Fourier Transform (ISTFT) method \cite{welch1967use} with Hanning window because of its less computational complexity than parametric synthesis approach \cite{mcdermott2011sound} \cite{slaney1995pattern} or example based synthesis method \cite{owens2016visually}. The inverse STFT (ISTFT) can regenerate the time-domain signal exactly from a given STFT without noise. We calculate ISTFT by inverting the STFT using the popular overlap-add (OLA) method \cite{ouelha2017efficient} for better denoising efficiency and synthesis quality as shown below:
\begin{equation}
y(t) = 1/{2\pi} \int_{-\infty}^{\infty} Y(\tau,w)e^{+jwt} d\tau dw
\end{equation}
For  the  phase  reconstruction from spectrogram, we apply the iterative Griffin-Lim algorithm \cite{griffin1984signal} for 16 iterations while performing ISTFT.  The proposed automatic Foley Generation Model details are summarized in Algorithm as follows:

\begin{algorithm}[H]
\caption{Automatic Foley Generation}
\textbf{Input: } Silent video frames ($v
_i$) and training audio tracks ($A_j$).\\
\textbf{Output: } Generated audio tracks ($Aud_{generated}$).

\begin{algorithmic}[1]
\FOR{$j \gets 1,2...N$}
\STATE{$S_j \gets Spectrogram (A_j)$}
\ENDFOR

\STATE{$Avg \gets Average(S)_{each class}$}

\IF{$Frame Sequence Network$}
\FOR{$i \gets 1,2...N$}
\STATE{$I_t \gets interpolate (v_i)$}
\STATE{$SP_t \gets spacetime (I_t)$}
\ENDFOR
\STATE{$V_t \gets concatenate (CNN(SP_t),CNN(I_1))$}
\STATE{$Model \gets FSLSTM(V_t)$}
\STATE{$Prob \gets Model.predict(V_{test})$}
\ENDIF

\IF{$Frame Relation Network$}
\FOR{$i \gets 1,2...N$}
\STATE{$I_t \gets interpolate (v_i)$}
\ENDFOR
\STATE{ $V_t \gets CNN(I_t)$}
\STATE{$Prob \gets TRN (V_t)$}
\ENDIF
\STATE{$S'\gets Prob + Avg$}
\STATE{$Aud_{generated} \gets ISTFT(S')$}

\end{algorithmic}
\end{algorithm}


\section{Experimental Result}

For our model evaluation, we create a video dataset particularly focusing on Foley tracks used for movies. We give a detail description of our dataset in subsection A. Next, we introduce the model parameters and implementation details in subsection B. Based on our dataset and models, we evaluate our generated sound both in qualitative and quantitative ways, described in subsection C and D respectively. Later in subsection E, we present a detailed ablation study of our methods and parameters. Finally, we discuss the results of four human evaluation survey questions on our synthesized sound in subsection F.

\subsection{The Automatic Foley Dataset (AFD)}
Our interest is to enable our Foley generation network to be trained with the exact natural sound produced in a particular movie scene. To do so, we need to train the system explicitly with the specific categories of audio-visual scenes that are closely related to manually generated Foley tracks for silent movie clips. We find different video datasets (e.g. GHD\cite{owens2016visually}, VEGAS\cite{zhou2017visual}, AudioSet\cite{gemmeke2017audio}, UCF101\cite{soomro2012ucf101}) for the sound generation task. However, none of these datasets are able to serve our requirements for various reasons. For example, the GHD dataset consists of videos of human actions (such as, hitting, scratching), that are mostly focused on a material recognition task. On the other hand, large number of videos are accumulated in the datasets like AudioSet, VEGAS, and UCF101, but most of them contain either background sounds, human speech or environmental noises. Foley tracks are generated and recorded in a denoising environment inside the studio. So we choose to create a dataset entirely focused on Foley sound tracks of movie events. Here we select 12 different categories of videos (that are frequently used for Foley generation) associated with clear sound tracks combining both indoor and outdoor scenes. We choose sound classes where we can record our own (e.g. cutting, footsteps, car passing, clock sound, breaking, etc.). For recording, we use a video camera along with a shotgun microphone system. Then we apply denoising algorithm used especially for outdoor recordings. For other popular Foley sound categories of movie clips such as, gunshots, horse running, waterfall, fire, rain, thunder sounds, we use YouTube and collect the most clear audio-video clips available with the least background noise. Altogether, our Automatic Foley Dataset (AFD) contains a total of 1000 videos from 12 different classes. The average duration of each video is 5 seconds. The twelve video classes and their associated data statistics are shown in Fig. 10 and 11 respectively.

\graphicspath{ {Figure} }
\begin{figure}[t]
\includegraphics[width=83mm]{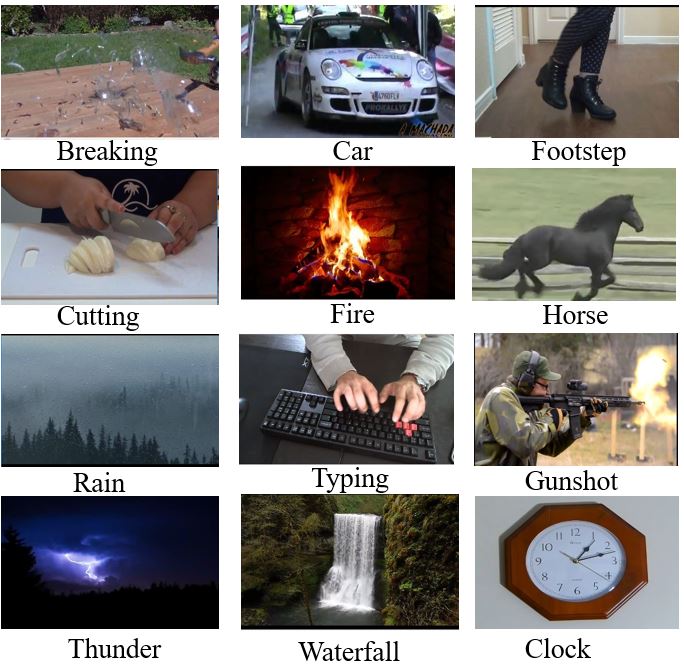}
\caption{The Automatic Foley Dataset (AFD) Video Classes: we chose 12 popular movie events where Foley effects are generally added in movie post-production studio.}
\end{figure}

\graphicspath{ {Figure} }
\begin{figure}[t]
\includegraphics[width=83mm]{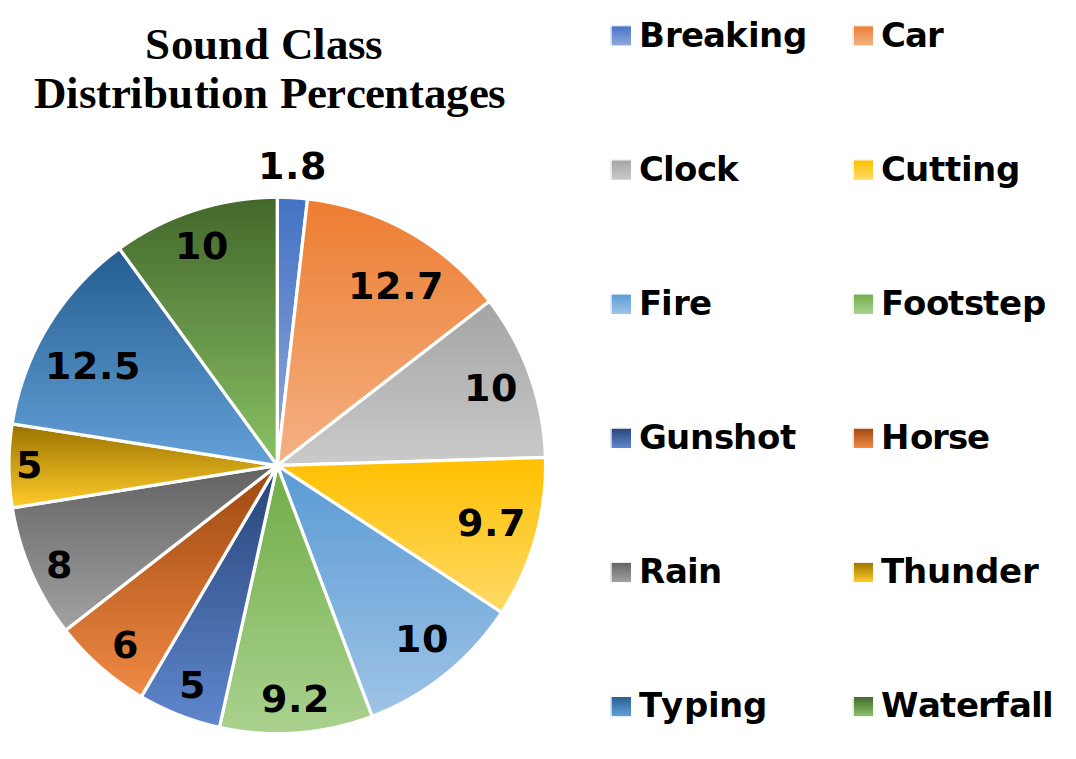}
\caption{The Automatic Foley Dataset (AFD) Distribution: we show the data distribution percentages of 12 video classes in the pie chart.}
\end{figure}

\subsection{Implementation Details}

We conduct the training separately, with our two approaches, using 80\% of the videos, representing all 12 classes in the dataset. Our audio sampling rate is 44 kHz and video frame rate is 190 FPS (after interpolation). For video interpolation technique, we use an m-interpolation filter in the ffmpeg video editing package from Python. In the first approach of sound classification, we get two (2048-D) image feature vectors from the output of $conv5$ layer of ResNet-50 network for both spacetime and first image frame from train videos. We concatenate these vectors to obtain the ultimate image feature vector of 4096-D that we pass to the FS-LSTM network as input. We apply 4 LSTM cells in the Fast layer of FS-LSTM because of its optimum performance in the classification task \cite{mujika2017fast}. At each LSTM cell, we set an initial value of 1 for the forget-bias. We use orthogonal matrices for all weight matrices. We apply layer normalization \cite{ba2016layer} separately to each gate. We use dropout\cite{srivastava2014dropout} to regularize the recurrent network. At every time step of FS-LSTM, we employ Zoneout \cite{krueger2016zoneout} in recurrent connections and a diverse dropout mask in non-recurrent connections \cite{zaremba2014recurrent}. During training, we use minibatch gradient descent with the Adam optimizer \cite{kingma2014adam}. Our minibatch size and learning rate are 0.001 and 128 respectively. In the second approach of sound classification, we keep the training hyper-parameters for the CNN architecture the same as earlier. In the TRN model, there are two layers of MLP (256 units in each) for $g_\theta$ and a single layer MLP (12 units) for $h_\phi$. The training for 100 epochs is completed in less than 18 hours on a Nvidia RTX 2080 Ti GPU. For testing our models, we choose 20\% of random videos for all 12 categories. To test the TRN model for early action recognition purpose, we select the first 25\% of frames in each video.
\subsection{Qualitative Evaluation}
\begin{figure*}[ht]
\centering
\begin{minipage}{\textwidth}
    \subfloat[Running horse]
    {
        \includegraphics[width=0.5\textwidth, height=7cm]{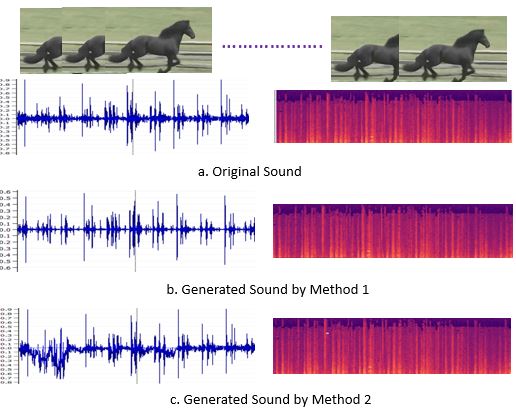}
        \label{fig:first_sub}
    }
    \subfloat[Typing]
    {
        \includegraphics[width=0.5\textwidth,height=7cm]{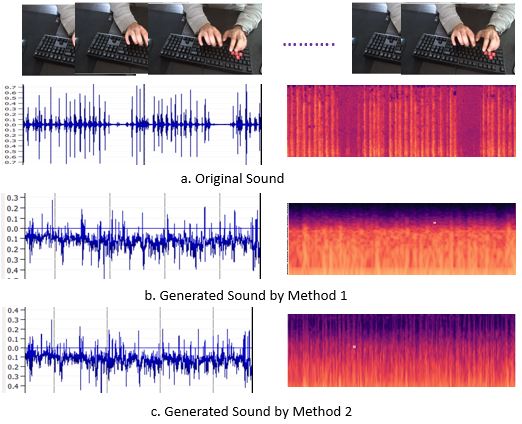}
        \label{fig:second_sub}
    }\\
    \subfloat[Ticking clock]
    {
        \includegraphics[width=0.5\textwidth, height=7cm]{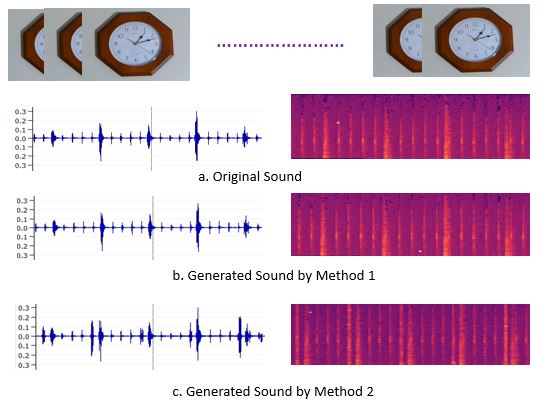}
        \label{fig:third_sub}
    }
    \subfloat[Waterfall]
    {
        \includegraphics[width=0.5\textwidth,height=7cm]{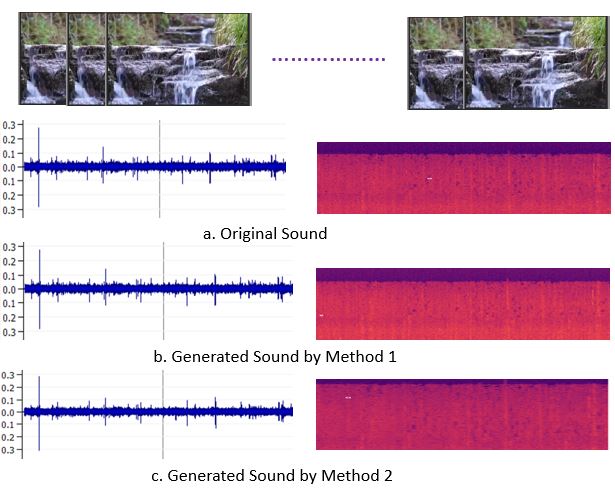}
        \label{fig:image2}
    }
    \caption{The waveform and spectral representation pair comparison between the original and generated sound.}
    \label{fig:image2}
\end{minipage}
\end{figure*}

\subsubsection{Waveform and Spectrogram Analysis}

For qualitative assessment, we present the synthesized sound waveform along with the corresponding spectrogram obtained from our two automatic Foley architectures with the same of original sound tracks in Fig.12. For some sound categories e.g. clock, rain, horse, we notice very similar waveform and spectrogram patterns of our generated sound to the original sound. However, results from model 1 matches more precisely with the ground truth patterns than model 2. We observe good alignment of synthesized and original sound for all sound categories that are less sensitive to meticulous timing (e.g. clock, fire, rain, waterfall). On the other hand, visual scenes containing random action variations with time (e.g. breaking things, cutting in kitchen, typing, gunshot in action sequences, lightning and thundering in sky) show few abrupt peaks and misalignment in the generated waveform. In video clips where sound sources are moving with distances (e.g. car racing, horse running, human footsteps), the sound amplitudes also vary with distance.

\subsubsection{Sound Quality Matrix Analysis}
Generally, the quality of a sound is assessed based on how well the sound conforms user’s expectations. We choose to evaluate to what extent our synthesized sound waves are correlated with their ground truth tracks. For each sound category, we calculate the average normalized cross correlation values between our generated and original audio signals to know how much similarity exists between them. We present the correlation values for models in Table I. We notice all positive cross-correlation values (above $0.5$) representing an expected analogy between the ground truth and our obtained result. Apart from the four most temporal action sensitive classes (e.g. breaking, cutting, footsteps, gunshots), model 1 provides higher correlation values than method 2. 

\definecolor{LightCyan}{rgb}{0.88,1,1}
\newcolumntype{P}[1]{>{\centering\arraybackslash}p{#1}}
\captionsetup[table]{name=TABLE,labelsep=newline,textfont=sc}
\begin{table}[h]
\begin{center}
\caption{\textcolor{Black}{Sound Quality Matrix Analysis Results: Average normalized cross-correlation value obtained from comparing the original and generated audio signals for model 1 and 2 in all sound classes}}
\label{tab:2}
\begin{tabular}{| p{20mm} | P{25mm} |  P{25mm} |}
\hline
 & \multicolumn{2}{c|}{\textbf{Avg. Normalized Correlation Value}} \\
 
\cline{2-3}
 \textbf{Sound Class} & \textbf{Model 1} & \textbf{Model 2}\\
\hline\hline
\rowcolor{LightCyan}
Break & 0.76 & 0.93   \\ \hline
Car & 0.68 & 0.65 \\ \hline
\rowcolor{LightCyan}
Clock & 0.92 & 0.73 \\ \hline
Cutting & 0.58 & 0.89\\ \hline
\rowcolor{LightCyan}
Fire & 0.88 & 0.72 \\ \hline
Footstep & 0.68 & 0.95 \\ \hline
\rowcolor{LightCyan}
Gunshot & 0.65 & 0.82 \\ \hline
Horse & 0.87 & 0.63 \\ \hline
\rowcolor{LightCyan}
Rain & 0.90 & 0.81 \\ \hline
Thunder & 0.71 & 0.58 \\ \hline
\rowcolor{LightCyan}
Typing & 0.69 & 0.60 \\ \hline
Waterfall & 0.86 & 0.69 \\ \hline
\rowcolor{Apricot}
\textbf{Average} & \textbf{0.77} & \textbf{0.75} \\
\hline
\end{tabular}
\end{center}
\end{table}

\subsubsection{Sound Retrieval Experiment}
The purpose of this qualitative task is to evaluate if semantic information of the sound class are present in our synthesized sound.  For this evaluation, we train a classifier model with the spectrograms of the original sound tracks from the AutoFoley training dataset. We use the ResNet-50 \cite{He_2016_CVPR} CNN architecture for our classifier network because of its excellent results on audio classification tasks in earlier work \cite{hershey2017cnn}. The major benefit of using 2D spectral representations resides in the summarizing capability of high dimensional signals into a compact representation through a spectrogram. After training, we feed the spectrograms of our synthesized sound clips obtained from both models to the trained classifier model for testing. We average the sound class prediction accuracy of all categories. The complete sound retrieval experiment model is shown in Fig.13. To evaluate our classifier performance and for better comparison, we also calculate the average prediction accuracy of the classifier network as $78.32\%$ by using the original spectrograms from our test dataset. Since prior works have not considered the movie sound effects domain, it is difficult to compare with their works directly. We present the prediction accuracy results of the most related sound generation models and our proposed models on the same retrieval task using classifier in Table II and III respectively. Both of our average prediction accuracies for this sound retrieval qualitative experiment show leading results above $63\%$ with our classifier for the generated sounds from AutoFoley Models. 
\begin{center}
\graphicspath{ {Figure} }
\begin{figure}[!h]
\fcolorbox{black}{white}{\includegraphics[width=83mm]{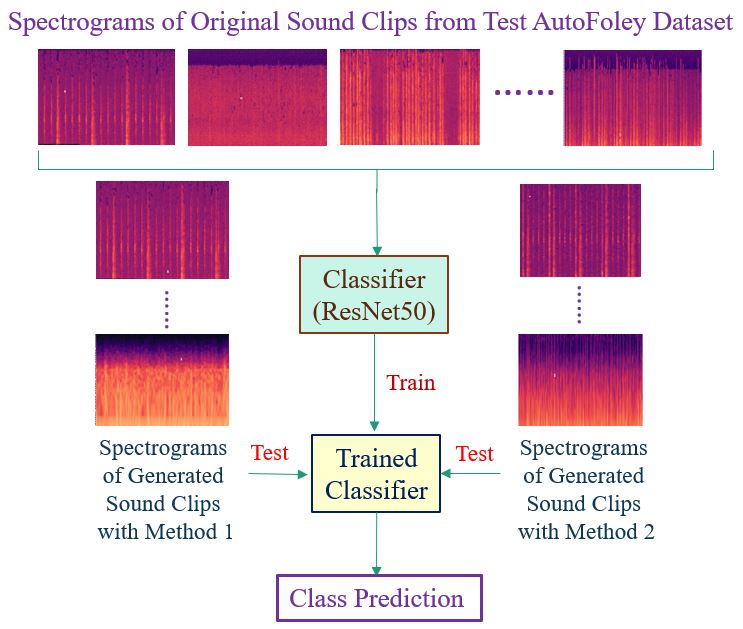}}
\caption{Sound Retrieval Experiment Model.}
\end{figure}
\end{center}

\captionsetup[table]{name=TABLE,labelsep=newline,textfont=sc}
\begin{table}[!h]
\begin{center}
\caption{Top 1 sound class prediction accuracy of existing models}
\begin{tabular}{|c|c|}
\hline
\textbf{Existing Model} & {\textbf{Avg. Acc.}} \\ \hline 
Owens et al.\cite{owens2016visually} (\textcolor{Blue}{ResNet + spectrogram + GHD dataset}) & 22.70\% \\\hline
POCAN \cite{zhangvisually} (\textcolor{RedViolet}{ResNet + spectrogram + GHD dataset}) & 36.32\% \\\hline
Zhuo \cite{zhou2017visual} (\textcolor{OliveGreen}{Flow method + VEGAS dataset}) & 45.47\% \\\hline
\end{tabular}
\end{center}
\end{table}

\captionsetup[table]{name=TABLE,labelsep=newline,textfont=sc}
\begin{table}[!h]
\begin{center}
\caption{Top 1 sound class prediction accuracy of proposed models with AutoFoley Dataset}
\begin{tabular}{|c|c|}
\hline
\textbf{ Proposed Model} & {\textbf{ Avg. Accuracy }} \\ \hline
AutoFoley (\textcolor{Purple}{Frame-Sequence Network}) & \textbf{65.79\%} \\\hline
AutoFoley (\textcolor{RoyalBlue}{Frame-Relation Network}) & \textbf{63.40\%}\\ \hline
AutoFoley (\textcolor{Mahogany}{Real sound tracks}) & 78.32\%\\ \hline
\end{tabular}
\end{center}
\end{table}
\subsection{Quantitative Evaluation}
It is a difficult task to obtain absolute numerical evaluation of generated sound waveforms. In this section we measure the audio-visual coherence ability of our models in predicting sound from given visual inputs. We also provide the calculated loss and accuracy details while training and testing our models.

\subsubsection{Sound Class Prediction}
To visualize the sound class prediction accuracy from video frames, we present two normalized confusion matrices for model 1 and 2 in Fig. 14. It is worth notifying that Model 1 correctly classifies a majority of the audio samples of the given test videos of all categories, except for breaking (as it is trained with least number of train samples). However, our second model can successfully identify the breaking class from test videos as the TRN network is trained to recognize the action present in a visual scene using the least amount of video frames. However, it miss-classifies some audios in the rain and waterfall cases as well as in cutting and footstep test videos. 

\begin{figure*}[]
\centering
\begin{minipage}{\textwidth}
    \subfloat[Model 1]
    {
        \includegraphics[width=0.48\textwidth, height=6cm]{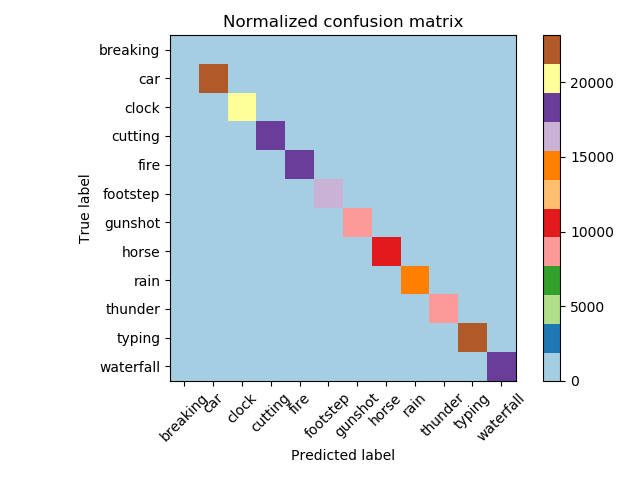}
        \label{fig:first_sub}
    }
    \subfloat[Model 2]
    {
        \includegraphics[width=0.48\textwidth, height=6cm]{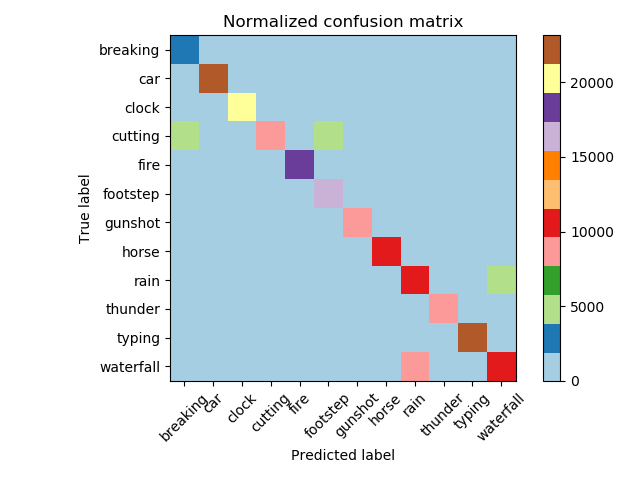}
        \label{fig:second_sub}
    }
\caption{Sound Class Prediction Result: Normalized Confusion Matrix of Model 1 and model 2.}
\label{fig:image2}
\end{minipage}
\end{figure*}

\subsubsection{Loss and Accuracy Calculation}
We calculate the average log loss and accuracy during training and testing of our models and display the results in Table IV. For both models, lower training losses are calculated as compared to the test case. Here, Model 1 gives a smaller log loss than Model 2 resulting in increased average accuracy throughout training and testing the network.


\begin{table}[!h]
\begin{center}
\renewcommand{\arraystretch}{1.3}
\caption{Loss and Accuracy Calculation Result}
\label{tab:example}
\begin{tabular}{|c|c|c|c|c|c|c|c|c|c|}
\hline
\textbf{Average} & \multicolumn{2}{c|}{\textbf{Train}} & \multicolumn{2}{ c |}{\textbf{Test}} \\ 
\cline{2-5}
& \textbf{Model 1} & \textbf{Model 2} & \textbf{Model 1} & \textbf{Model 2}\\
\hline \hline
\textcolor{Emerald}{Log Loss} & 0.002 & 0.035 & 0.166 & 0.194\\ \hline
\textcolor{Mulberry}{Accuracy} &  0.989 & 0.962 & 0.834 & 0.806\\ \hline
\end{tabular}
\end{center}
\end{table}

\subsection{Ablation Study}
In this section, we carry out and present supplementary experiments with the objective of validating the AutoFoley models described in the following subsections. 

\captionsetup[table]{name=TABLE,labelsep=newline,textfont=sc}
\begin{table*}[!h]
\begin{center}
\caption{\textcolor{Black}{Comparison of the average accuracy (\%) among the Ablation Models and our proposed method 1 (Frame Sequence Network) on AutoFoley Dataset.}}
\begin{tabular}{|c|c|c|}
\hline
\textbf{Model} & \textbf{Description} & \textbf{Average Accuracy (\%)} \\ \hline 
Ablation Model 1 & Direct  raw  image  frame  (no  SP)  +VGG19 + simple LSTM & 43.18\\
Ablation Model 2 & Direct raw image (no SP) + ResNet-50 + simple LSTM & 44.52\\
Ablation Model 3 & SP+ VGG19 + simple LSTM & 56.04\\
Ablation Model 4 & SP + VGG19 + FS-LSTM & 63.88\\
\textcolor{RoyalBlue}{Frame Sequence Network} & \textcolor{RoyalBlue}{SP + ResNet50 + FS-LSTM} & \textcolor{RoyalBlue}{65.79}\\
\hline
\end{tabular}
\end{center}
\end{table*}

\captionsetup[table]{name=TABLE,labelsep=newline,textfont=sc}
\begin{table*}[!h]
\begin{center}
\caption{\textcolor{Black}{Comparison of the average accuracy (\%) among the Ablation Models and our proposed method 2 (Frame Relation Network) on AutoFoley Dataset.}}
\begin{tabular}{|c|c|c|}
\hline
\textbf{Model} & \textbf{Description} & \textbf{Average Accuracy (\%)}\\
\hline
Ablation Model 1 & TRN (for 4 frame relations) + VGG19 (base CNN) & 41.64\\
Ablation Model 2 & TRN (for 4 frame relations) + ResNet-50 (base CNN) & 42.01\\
Ablation Model 3 & TRN (for 8 frame relations) +VGG19 (base CNN) & 60.98\\
Ablation  Model  4  &  TRN  (for  16  frame  relations)  +ResNet-50 (base CNN) & 64.27\\
\textcolor{Green}{Frame Relation Network} & \textcolor{Green}{TRN  (for  8  frame  relations)  +ResNet-50 (base CNN)} & \textcolor{Green}{63.40}\\
\hline
\end{tabular}
\end{center}
\end{table*}

\captionsetup[table]{name=TABLE,labelsep=newline,textfont=sc}
\begin{table*}[!h]
\begin{center}
\caption{\textcolor{Black}{Effect of Interpolation technique: Performance of the Frame-Sequence and Frame-Relation Networks  with  frame  replication  (k  factor)  instead  of  Interpolation. Comparing accuracy with proposed system with the Interpolation method}}
\begin{tabular}{|c|c|c|}
\hline
\textbf{Model} & \textbf{Description} & \textbf{Average Accuracy (\%)}\\
\hline
Ablation Model 1 & Frame Sequence Network using replicated video frame instead of Interpolation & 47.22\\
Ablation Model 2 & Frame Relation Network using replicated video frame instead of Interpolation & 43.83\\
\textcolor{RoyalBlue}{Proposed Model 1} & \textcolor{RoyalBlue}{Frame Sequence Network using Interpolation instead of replicated video frame} & \textcolor{RoyalBlue}{65.79}\\
\textcolor{Green}{Proposed Model 2} & \textcolor{Green}{Frame Relation Network using Interpolation instead of replicated video frame} & \textcolor{Green}{63.40}\\
\hline
\end{tabular}
\end{center}
\end{table*}

\subsubsection{Ablation Analysis with Frame-Sequence Network}
 
The proposed Frame-Sequence Network (described in Section III) contains three main steps: 1) the preprocessing of video frames by generating space time (SP) images, 2) image feature extraction with pretrained ResNet-50 CNN model, and 3) sound class prediction using a Fast-Slow LSTM model. To understand the significance of each component of this prediction network, we build the following alternative cascaded ablation models, train them with the AutoFoley dataset, then compute the prediction accuracy in each case and compare with the obtained result from our proposed Model 1 (Frame-Sequence Network):
\begin{itemize}
    \item \textbf{Ablation Model 1 }: Feed the network with direct raw image frame rather than creating space-time images. For classification we use VGG19 and a simple LSTM network as our CNN-RNN model.
    \item \textbf{Ablation Model 2} : Apply direct raw image without using space-time images. Here we add ResNet-50 as CNN model with a simple LSTM network.
    \item \textbf{Ablation Model 3} : Perform the classification using VGG19 and a simple LSTM that are trained on space-time images rather than raw images.
    \item \textbf{Ablation Model 4} : Replace the simple LSTM network with FS-LSTM. Other steps are followed as same as Ablation Model 3.
\end{itemize}

The Table V shows the performance comparisons among these ablation models and our proposed Frame Sequence model. There is a noticeable accuracy degradation for the models that are using direct raw images as video frame input to the CNN. This is because they often miss useful color and motion features from consecutive video frames. The use of a simple LSTM design reduces the computation time, however, it provides higher class prediction errors. Though accuracy computed from the last ablation model (including SP, VGG19, FS-LSTM) is close to the proposed Frame-Sequence Network result, our model outperforms all other cascaded models.       

\subsubsection{Ablation Analysis with Frame-Relation Network}
We now assess the effect of the parameter $Q$ in the multi-scale TRN. To assess $Q$, we train and test the ablation models for different $Q$ values and observe their performances. To evaluate the effects of parameters and models used in our second proposed model (Frame-Relation Network), we train and test the following ablation models and compare their performance.  
\begin{itemize}
    \item \textbf{Ablation Model 1 }: In this model we apply musti-scale TRN, adding up to 4 frame relations of video while using VGG19 as the base CNN.  
    \item \textbf{Ablation Model 2} : Change the base CNN VGG19 with ResNet-50. Remaining algorithm remains same as the previous ablation model.  
    \item \textbf{Ablation Model 3} : We add up to 8 frame relations in the multi-scale TRN and use VGG19 again as base CNN.
    \item \textbf{Ablation Model 4} : Raise the frame relation number to 16 frames in the multi-scale TRN and apply ResNet-50 for CNN.  
    \item \textbf{Proposed Model 2} : In our proposed Frame-Relation Network (as described in earlier sections) we apply a multi-scale temporal relational network computing the relation over time up to 8 video frames ($Q$= 8) using ResNet-50 as the base CNN.  
\end{itemize}

The detailed comparative results are shown in Table VI. Here, we observe lower accuracies in cases of the 4 frame relation TRN. The accuracy improves with the increment of $Q$. However, it shows a noticeable computational time increment after setting $Q$ value to 16 compared to 8, resulting a marginal accuracy improvement. Hence, in our proposed network, we do not further increase the $Q$ value above 8. We also find that, ResNet-50 outperforms VGG19 model as the base CNN. 

\subsubsection{Effect of Interpolation Technique}

To maintain the same length of visual and sound sequences while mapping, earlier works \cite{owens2016visually, zhou2017visual} adopted the technique of duplicating the video frames. In the flow-based method \cite{zhou2017visual}, pre-computed optical flow between video frames is fed to the temporal ConvNets to explicitly capture the motion signal. But adding an optical flow-based deep feature to the visual encoder requires global optimization, thus it is difficult to implement. Also, flow-based methods tend to cause visual artifacts with brightness variations, as they are sensitive to parameter settings. In contrast to the optical flow method, and replication of the same video frame for multiple times to map with a single sound sample, we use an interpolation technique to leverage intermediate video frames in order to obtain smooth motion information of each video. Interpolation technique can efficiently deal with large data volumes for the higher frame rate videos in movies. Standard optical flow techniques often become less efficient for interpolating this type of densely sampled, large-scale data. We exploit the video interpolation technique before feeding the input frames to deep neural networks in order to remove the difference between the video frame rate and the audio sampling rate while mapping audio-video features. Again, this pre-processing method allows precise capture of the action features in fast moving video frames. The interpolation technique removes the limitations faced in the replication method proposed  in earlier works \cite{zhou2017visual, owens2016visually}. In Table VII, we show the comparison of overall accuracies after applying these two alternative pre-processing methods to the AutoFoley dataset.

\subsubsection{Effect of Sound Generation tool}
  In prior work, \cite{zhou2017visual}, a three tier SampleRNN is used as the sound generator for creating audio samples one at a time at a sampling rate of 16KHz. As a result, this auto-regressive model in SampleRNN undergoes inference times and prohibitive training. Another drawback of this sound synthesizer is that it reduces the resolution of the predictions in order to lessen the computational cost. Hence, for sound generation, we use the Inverse Short Time Fourier Transform (ISTFT) method, which is comparatively simpler and has less computational complexity. The generated sound wave is calculated as the Inverse Short Time Fourier Transform (ISTFT) of the predicted spectrogram, which is the sum of predicted regression parameters and a base sample corresponding to the predicted sound class.

\subsection{Human Evaluation}
To evaluate the quality of our synthesized sounds, we conduct a survey among local college students. In the survey, the students are presented a video with two audio samples, the original sound and the synthesised sound. We then ask students to select the option they prefer using four questions:

\begin{enumerate}
    \item Select the original (more realistic) sample.
    \item Select the most suitable sample (most reasonable audio initiated from the video clip).
    \item Select the sample with minimum noise. 
    \item Select the most synchronized sample.
\end{enumerate}

We assess the performance of our produced sound tracks for each of the categories. We evaluate both of our approaches through a comparison task with the ground truth in the first query. From here, we like to gauge how realistic are our synthesized Foley tracks. To evaluate this, we create a survey of a video option containing its original sound track and the same video in another option containing the generated sound tracks, for each of our AutoFoley categories. We observe when people make a wrong choice between the generated sound and the original one. In this survey 73.71\% of the respondents chose the synthesised sound over the original sound with our first model, and 65.95\% with the second model.  

The remaining three qualitative questions are used to compare the performance between the two models. In these queries, we set the same two videos of all classes in two options associated with synthesized sound tracks from Models 1 and 2 respectively. We evaluate which method is preferred by respondents after observing the audio-video pairs. The survey results show that Model 2 (Frame Relation Network + ISTFT) outperforms Model 1 (Frame Sequence Network + ISTFT) for visual scenes associated with random action changes (i.e. breaking, cutting, footsteps, gunshot sound classes). For the rest of the categories, respondents chose the Model 1 generated sound tracks over Model 2. The detailed human evaluation results (selection percentages of survey queries) for each individual class are presented in Tables VIII and IX for Models 1 and 2 separately. 
\begin{table}[!h]
\begin{center}
\caption{\textcolor{Black}{Human Evaluation Results: Selection percentage of each sound category for the first and second human survey questions}}
\begin{tabular}{| c | c | c | c | c |}
\hline
\textbf{Class} & \multicolumn{2}{ c |}{\textbf{Query 1}} & \multicolumn{2}{ c |}{\textbf{Query 2}} \\ 
\cline{2-5}
& \textbf{Method 1} & \textbf{Method 2} & \textbf{Method 1} & \textbf{Method 2}\\ \hline
\rowcolor{LightCyan}
Break & 52.40\% & 56.10\% & 32.30\% & 67.70\%  \\ \hline
Car & 71.53\% & 65.40\% & 55.76\% & 44.24\% \\ \hline
\rowcolor{LightCyan}
Clock & 90.91\% & 73.80\% & 70.90\% & 29.10\%\\ \hline
Cutting & 50.17\% & 45.35\% & 62.89\% & 37.11\%\\ \hline
\rowcolor{LightCyan}
Fire & 85.43\% & 75.40\% & 57.70\% & 42.30\% \\ \hline
Footstep & 61.72\% & 50.57\% & 72.13\% & 27.87\% \\ \hline
\rowcolor{LightCyan}
Gunshot & 68.33\% & 61.84\% & 64.60\% & 35.40\%\\ \hline
Horse & 89.38\% & 78.20\% & 53.80\% & 46.20\% \\ \hline
\rowcolor{LightCyan}
Rain & 88.62\% & 75.80\% & 50.00\% & 50.00\% \\ \hline
Thunder & 76.25\% & 72.17\% & 59.35\% & 40.65\%\\ \hline
\rowcolor{LightCyan}
Typing & 64.27\% & 62.39\% & 66.20\% & 33.80\% \\ \hline
Waterfall & 85.55\% & 74.40\% & 66.78\% & 33.22\%\\ \hline
\rowcolor{Apricot}
\textbf{Average} & \textbf{73.71\%} & \textbf{65.95\%} & \textbf{59.37\%} & \textbf{40.63\%} \\ \hline
\end{tabular}
\end{center}
\end{table}

\begin{table}[!h]
\begin{center}
\caption{\textcolor{Black}{Human Evaluation Results: Selection percentage of each sound category for the third and fourth human survey questions}}
\begin{tabular}{| c | c | c | c | c |}
\hline
\textbf{Class} & \multicolumn{2}{ c |}{\textbf{Query 3}} & \multicolumn{2}{ c |}{\textbf{Query 4}} \\ 
\cline{2-5}
& \textbf{Method 1} & \textbf{Method 2} & \textbf{Method 1} & \textbf{Method 2}\\ \hline
\rowcolor{LightCyan}
Break & 29.17\% & 70.83\% & 29.00\% & 71.00\% \\ \hline
Car & 36.02\% & 63.08\% & 57.32\% & 42.68\%\\ \hline
\rowcolor{LightCyan}
Clock & 57.69\% & 42.30\% & 61.50\% & 38.50\%\\ \hline
Cutting & 62.50\% & 37.50\% & 55.41\% & 44.59\%\\ \hline
\rowcolor{LightCyan}
Fire & 34.61\% & 65.38\% & 53.80\% & 46.20\%\\ \hline
Footstep & 65.37\% & 34.62\% & 68.55\% & 31.45\%\\ \hline
\rowcolor{LightCyan}
Gunshot & 75.81\% & 24.19\% & 74.60\% & 25.40\%\\ \hline
Horse & 36.18\% & 63.82\% & 53.80\% & 46.20\%\\ \hline
\rowcolor{LightCyan}
Rain & 33.33\% & 66.67\% & 55.80\% & 44.20\%\\ \hline
Thunder & 76.92\% & 23.08\% & 51.62\% & 48.38\%\\ \hline
\rowcolor{LightCyan}
Typing & 67.24\% & 32.76\% & 68.50\% & 31.50\%\\ \hline
Waterfall & 51.85\% & 48.15\% & 52.43\% & 47.57\%\\ \hline
\rowcolor{Apricot}
\textbf{Average} & \textbf{52.23\%} & \textbf{47.70\%} & \textbf{52.86\%} & \textbf{43.14\%}\\ \hline
\end{tabular}
\end{center}
\end{table}

\section{Conclusion}
In this paper, we address a novel problem of adding Foley effect sound tracks to video clips of movies by using an efficient deep learning solution. Here, we have proposed two deep neural models, Frame Sequence and Frame Relation Network (associated with a less complex sound synthesis approach). These models are trained to predict the sound features from only visual inputs and artificially synthesize the required sound track. We have also introduced a nascent dataset for this particular task that contains audio-video pairs of the most popular Foley scene categories. Our models show increased computational efficiency in learning the intricate transition relations and temporal dependencies of visual inputs from a low number of video frames. Though prior works do not directly match with our work, we conduct extensive qualitative, numerical and ablation analysis to demonstrate the usefulness of the proposed models and tools. We have achieved higher accuracy in sound retrieval experiment results (over 63\%) for both prediction models than state-of-the-art researches, i.e. sound generation from visuals. Lastly, our human survey result shows greater than 73\% of respondents considered our generated sound as original. 

Since the task of adding automatic Foley to silent video with deep learning is an interesting and novel task, the next steps in this research are to expand on the training dataset, allowing for the generated sound output to more closely approximate the original sound. The computational efficiency of the proposed model will also be improved, with the goal of being able to process live video in real-time. The time synchronization problem will also be examined and optimized in future research.

\bibliographystyle{IEEEtran}
\bibliography{AFoley.bib}{}
\begin{IEEEbiography}[{\includegraphics[width=1in,height=1.25in,clip,keepaspectratio]{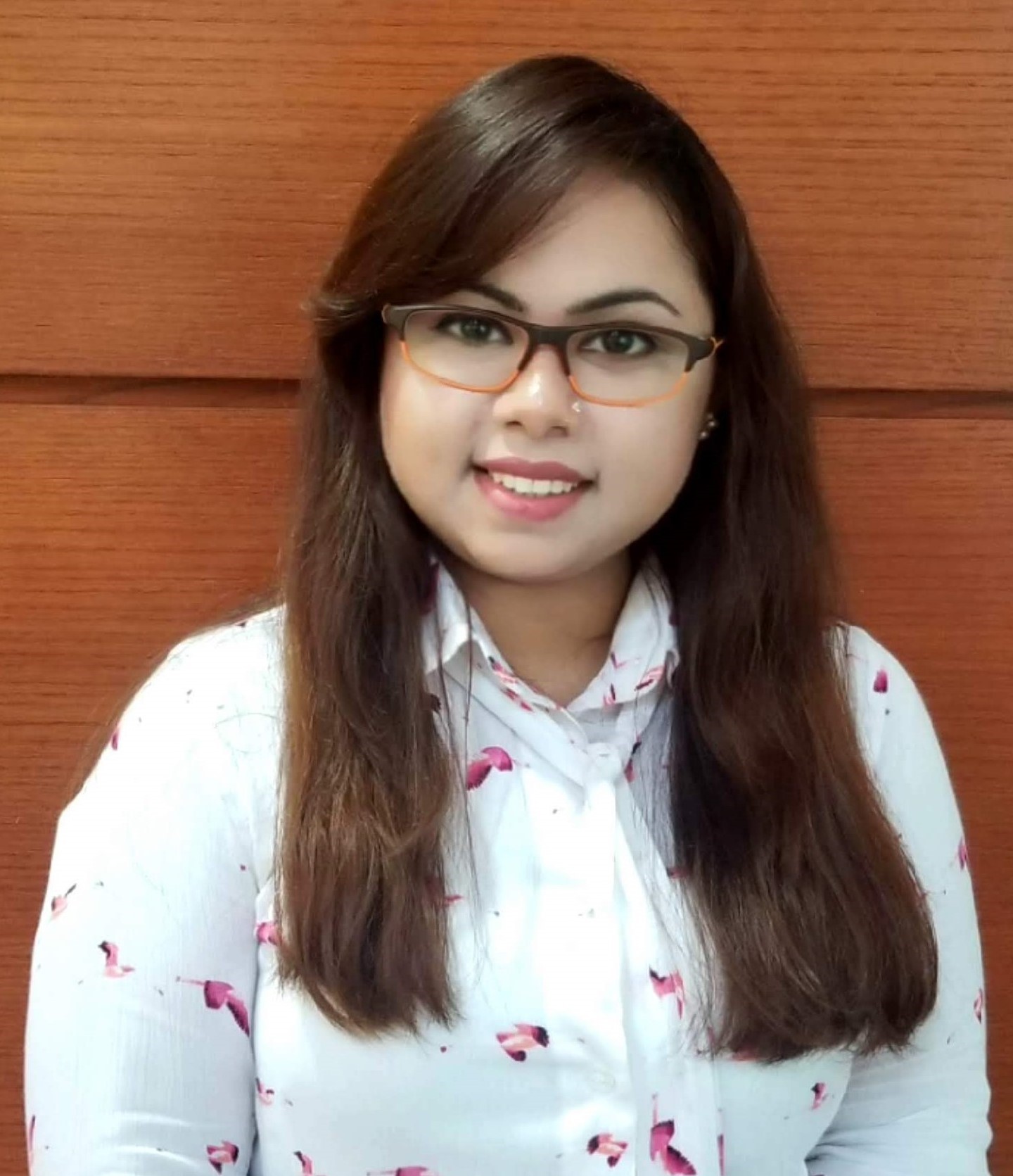}}]{Sanchita Ghose} 
received the bachelor’s degree in Electrical and Electronic Engineering from Ahsanullah University of Science and Technology (AUST), Dhaka, Bangladesh
in 2013. She is currently working toward the doctoral degree with the Cloud Lab for Engineering Application and Research (CLEAR), in the University of Texas at San Antonio, Texas, USA. She is a member of  the  Open  Cloud  Institute (OCI), IEEE, and Society of Women Engineers (SWE). Her current research interest includes developing deep learning algorithm for multi-modal learning and cross modal retrieval applications, focusing on computer vision, action recognition, sound synthesis and video processing.
\end{IEEEbiography}

\begin{IEEEbiography}[{\includegraphics[width=1in,height=1.25in,clip,keepaspectratio]{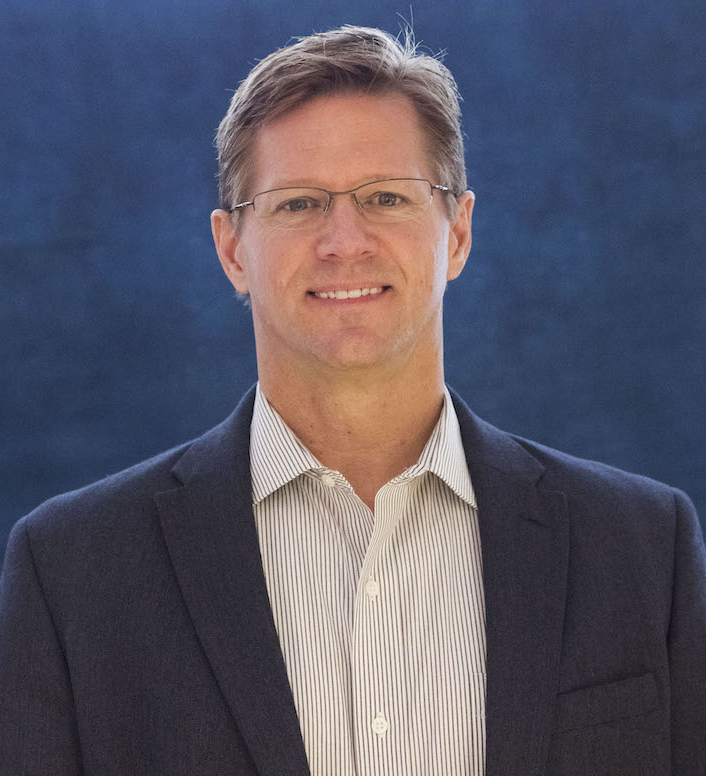}}]{John Jeffrey Prevost (S'06, M'13)}
received his first B.S. degree from Texas A\&M in Economics in 1990. He received his second B.S. degree in Electrical Engineering from the University of Texas at San Antonio (UTSA), where he graduated Magna Cum Laude in December 2009. In 2012 he received his M.S. degree in Electrical Engineering, also from UTSA along the way to earning his Ph.D. in Electrical Engineering in December, 2013. His current academic appointment is Assistant Professor in the department of Electrical and Computer Engineering at UTSA. In 2015, he co-founded and became the Chief Research Officer and Assistant Director of the Open Cloud Institute. Prior to his academic appointment he has served as Director of Product Development, Director of Information Systems and Chief Technical Officer for various technical firms. He remains an active consultant in areas of complex systems and cloud computing and maintains strong ties with industry leaders. His current research interests include energy aware cloud optimization, cloud controlled robotics, cloud based communications, and quantum cloud computing.
\end{IEEEbiography}
\end{document}